\documentclass[twocolumn,superscriptaddress,aps]{revtex4}
\usepackage{amsmath}
\usepackage{amssymb}
\usepackage{graphicx}
\usepackage{color}
\usepackage[colorlinks=true,linkcolor=blue,citecolor=blue]{hyperref}

\newcommand{\be}{\begin{equation}}
\newcommand{\ee}{\end{equation}}
\newcommand{\bea}{\begin{eqnarray}}
\newcommand{\eea}{\end{eqnarray}}
\newcommand{\bse}{\begin{subequations}}
\newcommand{\ese}{\end{subequations}}
\newcommand{\parent}{${\rm SrMnSb_2}$}
\newcommand{\Ksub}{${\rm (Sr_{0.97}K_{0.03})MnSb_2}$}

\newcommand{\Tn}{$T_{\textrm N}$}
\newcommand{\Tc}{$T_{\textrm c}$}

\begin{document}

\title{Hole Doping and Antiferromagnetic Correlations above the N{\'e}el temperature of the Topological Semimetal (Sr$_{1-x}$K$_x$)MnSb$_2$ }
\author{Yong Liu}
\thanks{Present address: Institute of High Energy Physics, Chinese Academy of Sciences, Beijing 100049, China and Dongguan Neutron Science Center, Dongguan 523803, China.} 
\author{Farhan Islam}
\author{ Kevin W. Dennis}
\affiliation{Division of Materials Sciences and Engineering, Ames Laboratory, U.S. DOE, Ames, Iowa 50011, USA}
\author{Wei Tian}
\affiliation{Neutron Scattering Division, Oak Ridge National Laboratory, Oak Ridge, Tennessee 37831, USA}
\author{Benjamin G. Ueland}
\author{Robert J. McQueeney}
\author{David Vaknin}
\email{vaknin@ameslab.gov}

\affiliation{Division of Materials Sciences and Engineering, Ames Laboratory, U.S. DOE, Ames, Iowa 50011, USA}
\affiliation{Department of Physics and Astronomy, Iowa State University, Ames, Iowa 50011, USA}

\date{\today}

\begin{abstract}

Neutron diffraction and magnetic susceptibility studies of orthorhombic single crystal  {\Ksub}  confirm the three dimensional (3D) C-type antiferromagnetic (AFM) ordering of the Mn$^{2+}$ moments at $T_{\rm N}=305 \pm 3$ K which is slightly higher than that of the parent  SrMnSb$_2$ with $T_{\rm N}=297 \pm 3$ K.  Susceptibility measurements of the K-doped and parent crystals  above $T_{\rm N}$ are characteristic of  2D AFM systems. This is consistent with  high temperature neutron diffraction of the parent compound that display persisting 2D AFM correlations well above  $T_{\rm N}$ to at least $\sim 560$ K with no evidence of a ferromagnetic phase.  Analysis of the de Haas van Alphen magnetic oscillations of the K-doped crystal  is consistent with  hole doping. 
\end{abstract}

\maketitle

\section{Introduction}
The magnetic semimetals $A$Mn$Pn_2$ ($A =$ Sr, Ba, Ca; $Pn =$ Sb and Bi) have attracted intense interest as they have the potential to exhibit  topological  electronic behaviors with the key  signature of Dirac fermions based on first-principles calculation, angle-resolved photoelectron spectroscopy (ARPES), and quantum oscillations while also ordering antiferromagnetically\cite{Farhan2014,Lee2013,Park2011,Wang2012,Wang2012a,Wang2012b}. These  compounds consist of alternate stacking of $Pn$-Mn-$Pn$ layers that are separated by $A$-$Pn$-$A$ layers. In the $A$-$Pn$-$A$ layer, $Pn$ atoms are arranged in a square, or a distorted square lattice, that are predicted to host Dirac fermions with adequate spin-orbit coupling\cite{Farhan2014}. Indeed, theoretical calculations indicate that the electronic bands associated with the $A$-$Pn$-$A$ layers are closer to the Fermi level than those arising from the $Pn$Mn$Pn$ layers, and display a linear energy-momentum dispersion similar to that in graphene, which is a key signature of the presence of Dirac fermions \cite{Lee2013}. 

Dirac Fermions in $A$Mn$Pn_2$  seem to display complex properties due to their interplay with the magnetism in the Mn planes leading to novel magneto-topological states. For example, it has been reported that Mn- and Sr-deficient Sr$_{1-y}$Mn$_{1-z}$Sb$_2$ with $y \sim 0.08$ and $z \sim 0.02$ is a Weyl semimetal that emerges by subjecting the Dirac fermions to a broken time-reversal symmetry via ferromagnetic (FM) ordering\cite{Liu2017}. Specifically, this system is reported to undergo a FM transition below $T_{\rm C} \simeq 565$ K, and then an AFM transition below  $T_{\rm N} = 304$ K with a uniform canted moment that retains  a weak net FM moment \cite{Liu2017}.   We note that we have recently found in nearly stoichiometric  SrMn$_2$Sb$_2$ that weak ferromagnetism is tied to impurities on the sample surface\cite{Liu2019}.

SbMnSb layers in $A$MnSb$_2$ (referred to as Mn-112 compounds) are common to a large family of layered pnictides such as, $A$Mn$_2Pn_2$ (referred to as Mn-122 compounds) and $R$Mn$Pn$O ($R =$ La, Ce, Pr ...; 1111 compounds) that all settle into a G- or C-type AFM ground states corresponding to N{\`e}el or checkerboard spin ordering within the square layer and either AFM or FM coupling between layers, respectively. \cite{An2009,Singh2009,Singh2009b,Zhang2015,McGuire2016,Zhang2016,Liu2018}.  It should be  noted that the AFM is highly anisotropic as the inplane exchange coupling among nearest neighbors (NN) Mn$^{2+}$ moments in the Mn$Pn$ planes ($J_{\rm 2D}$) is much stronger than the weaker inter-plane coupling $J_{\rm i}$. As is usually observed in these and other layered systems \cite{Vaknin1989,Sangeetha2016}, two-dimensional (2D) AFM correlations develop at temperatures that are significantly higher than the three-dimensional (3D) N{\'e}el temperature ($T_{\rm N}$) \cite{Curely1998}.  Consequently, magnetic susceptibility  versus temperature [$\chi(T)$] measurements show a very broad peak at temperatures that are much higher than $T_{\rm N}$ at an onset of emerging 2D AFM correlations \cite{Curely1998}. In particular, for powder samples, $\chi(T)$ does not  show strong features at $T_{\rm N}$ and only weak anomalies in ${\rm d}\chi(T)/{\rm d}T$ that reveal the transition to the 3D AFM phase.

Uniform moment canting in the AFM phase that produces a net FM moment, as  has been reported for Sr$_{1-y}$Mn$_{1-z}$Sb$_2$ with $y \leq 0.1$\cite{Liu2017}, is one way to break  time-reversal symmetry. Another approach to achieving FM, is inspired by recent studies that show that hole doping by substituting sufficient Ba with K (or Rb) as in (Ba$_{1-x}$K$_x$)Mn$_2$As$_2$  the localized antiferromagnetism on the Mn site is preserved and coexists with itinerant ferromagnetism associated with As bands.\cite{Pandey2013,Ueland2015}.  Motivated by these reports, we studied similar substitutions in the Mn-112 systems that may provide an alternative route to inducing itinerant FM associated with the electronic band of Sb. 

Here, we report on the magnetic properties of single crystals of SrMnSb$_2$ and (Sr$_{0.97}$K$_{0.03}$)MnSb$_2$ using neutron diffraction technique in conjunction with various magnetization measurements. Although we find that the system can be hole doped by K substitution of the Sr site,  we also find that our approach to  substitution is limited to values that are not be sufficient to induce the desired FM state that is required for the Weyl state to exist. 

\section{Experimental Details}
Plate-like SrMnSb$_2$ single crystals were grown by a self-flux method  (more details can be found in Ref. \cite{Liu2019}). K substituted samples were prepared from a starting molar ratio of Sr:K:Mn:Sb=$1-x:2x:1:4$. The K content was doubled to avoid any loss during the crystal growth as it is an active and volatile element. We tried nominal $x=0.1, 0.2$ and $0.5$ compositions. For $x=0.1$ and $0.2$, the crystals were found to have an actual K concentration of $x=0.03$, as determined by Energy-dispersive X-ray spectroscopy (EDS) measurements. For the $x=0.5$ growth, no crystals were obtained. This suggests that there is an extremely limited solubility of K in  {\parent}. 

\begin{figure}
\centering
\includegraphics[width=0.8\linewidth]{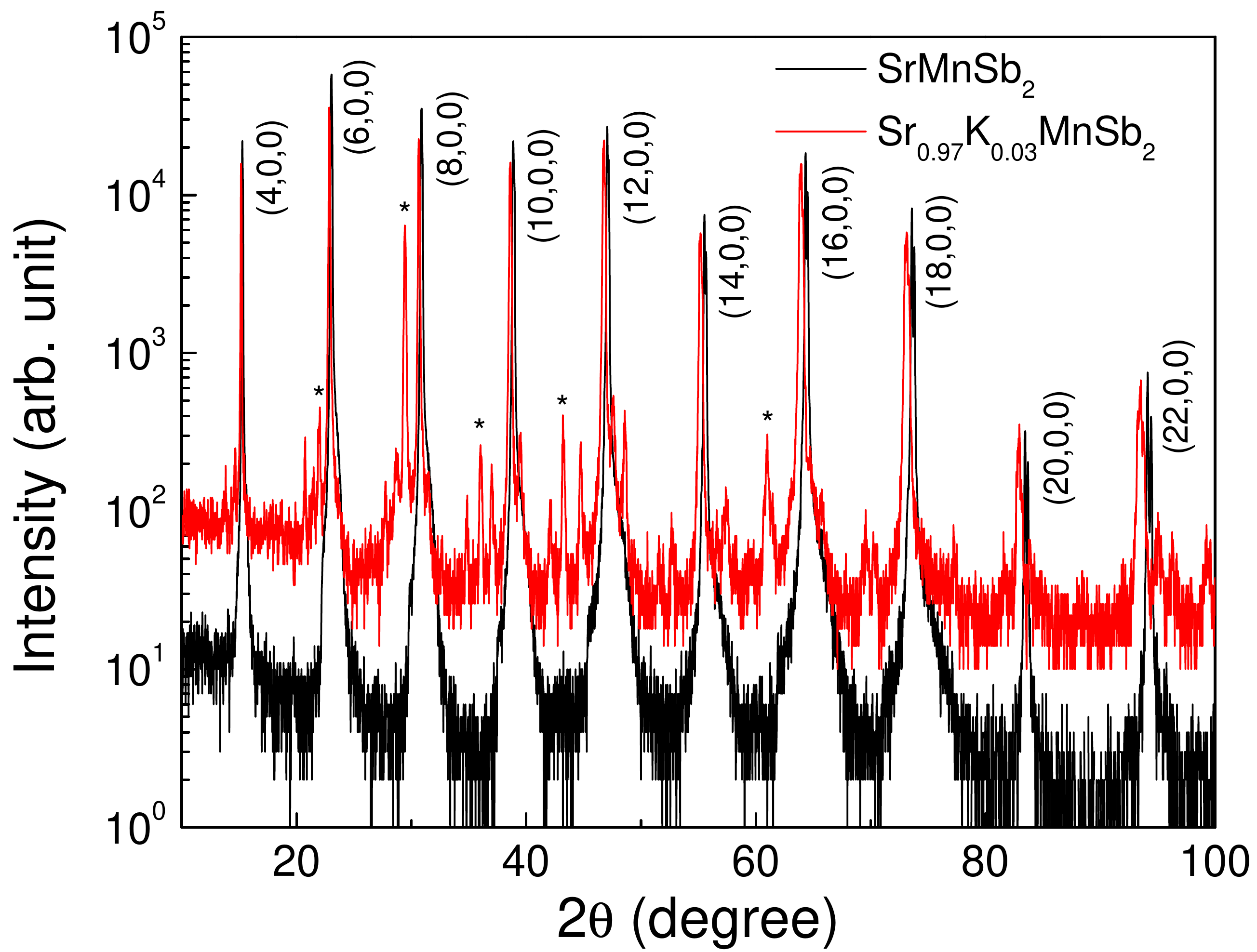}
\caption{(Color online) XRD patterns displaying the ($H$,0,0) reflections up to the 22nd order for both crystals by using Cu-K$\alpha$ radiation, shown on log scale. Weak extra peaks for (Sr$_{0.97}$K$_{0.03}$)MnSb$_2$ crystals,  indicated by the asterisk, could not be matched with known impurity phases unequivocally.} 
\label{Fig:XRD}
\end{figure}

X-ray diffraction (XRD) measurements were performed on a Bruker D8 Advance Powder Diffractometer using Cu K$\alpha$ radiation. The single crystal was fixed to a sample stage by using clay. The surface of the plate-like crystal was aligned with the surface of the edge of the sample stage by using a glass slide. Figure \ref{Fig:XRD} shows XRD patterns of SrMnSb$_2$ and (Sr$_{0.97}$K$_{0.03}$)MnSb$_2$ single crystals with their ($H$00) in the scattering plane. A set of ($H$00) diffraction peaks to the 22nd order is observed for both samples, from which we extract  the  lattice parameter $a$ (which is perpendicular to the layers in the $Pmna$ space group) and  determine as 23.051(5) and 23.27(5) {\AA} for SrMnSb$_2$ and (Sr$_{0.97}$K$_{0.03}$)MnSb$_2$, respectively. The slight increase in the $a$-spacing is expected as the atomic radius of K is slightly larger than that of Sr.

Magnetization measurements were performed by using a Physical Property Measurement System (PPMS, Quantum Design) equipped with a vibrating sample magnetometer (VSM). For the temperature dependent magnetization measurements, the sample was cooled down to the desired temperature without magnetic field or with an application of magnetic field, termed as ZFC and FC, respectively. The temperature dependent magnetization data were then collected upon warming at 2 K/min and at a fixed field. The magnetic fields $H$ were applied parallel to the plate ({\bf H}$\perp ${\bf a}) and perpendicular to the plate ({\bf H}$||${\bf a}). 

Single crystal neutron diffraction experiments were carried out on the HB-1A triple axis spectrometer located at the High Flux Isotope Reactor  at Oak Ridge National Laboratory. The HB-1A spectrometer operates with a fixed incident energy of $E_i =14.64$ meV using a double pyrolytic graphite (PG) monochromator system. PG filters were placed before and after the second monochromator to reduce higher order contamination in the incident beam achieving a ratio $\frac{I_\lambda}{2}:I_\lambda: \approx 10^{-4}$. The  {\Ksub} and {\parent} crystals were mounted with the ($HK0$)  in the scattering plane (40'-40'-S-40'-80' collimation.  The {\parent}  was also mounted with ($0KL$) in the scattering plane with a tighter resolution (40'-10'-S-10'-80') to resolve the in-plane orthorhombic distrortion. We note that some results on the parent SrMnSb$_2$ are reproduced from Ref.\cite{Liu2019} for comparison with the K-doped sample and some are new. 

\section{Results and Discussion}

\begin{figure}
\includegraphics[width=3. in]{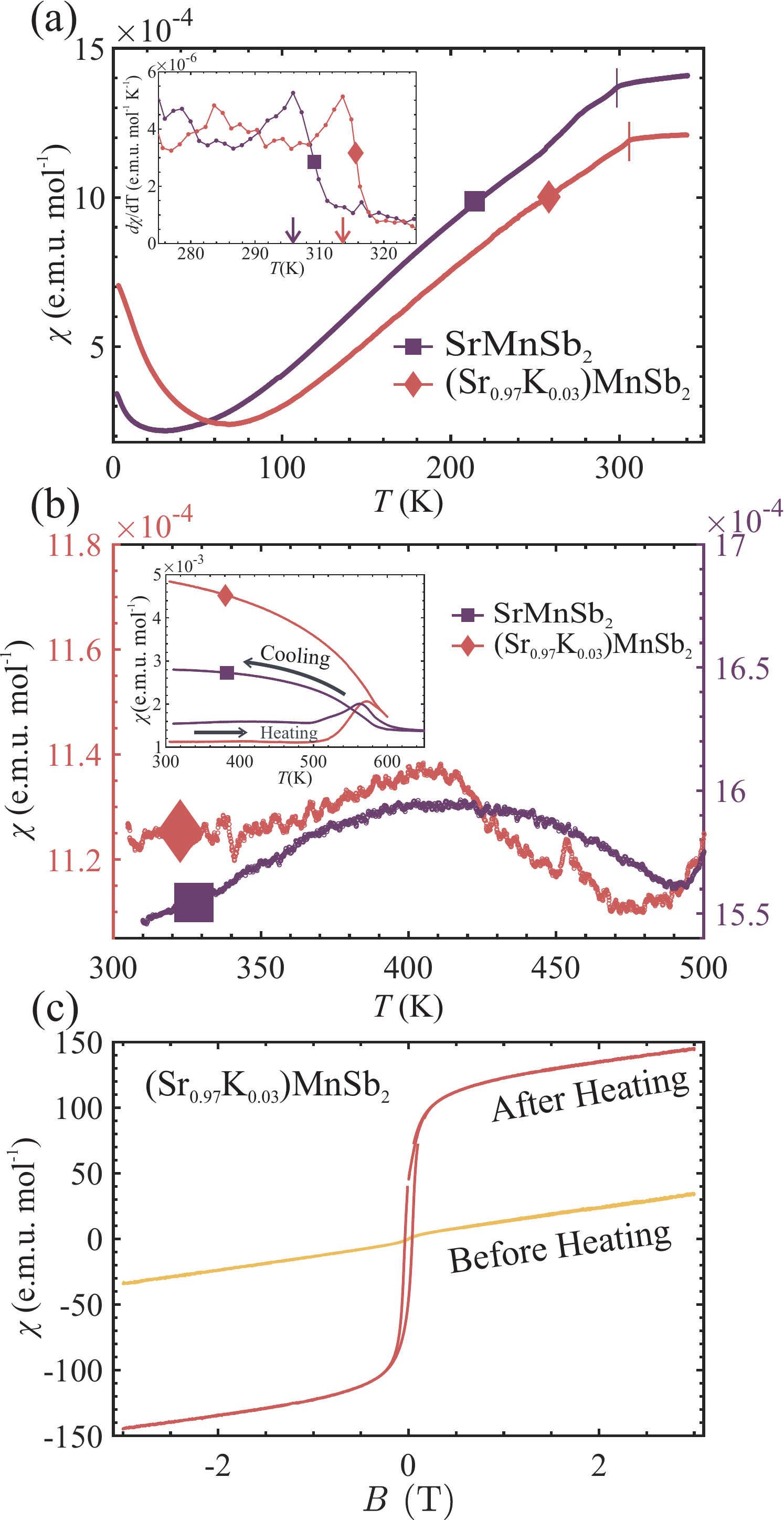}
\caption{(Color online) (a) Zero-field-cooled (ZFC)  magnetic susceptibility, $\chi\equiv M/H$, of single crystals {\parent} and {\Ksub} with applied magnetic field along the $a$-axis ($\chi_a$) for SrMnSb$_2$ and (Sr$_{0.97}$K$_{0.03}$)MnSb$_2$ indicating a slight increase in {\Tn} upon K-substitution. The inset shows d$\chi (T)$/d$T$ versus $T$ with a peaks at {\Tn} that coincide with neutron diffraction extracted {\Tn}'s. The K-substituted single crystal exhibits a slightly higher {\Tn} than the parent compound. (b) ZFC susceptibility  at high temperatures (above {\Tn}) for the parent and the doped crystals with magnetic field aligned in the $bc$-plane.  The increase in susceptibility above the  {\Tn}  and the broad peak are common signatures  of 2D magnetic systems that develop large correlations prior to the 3D ordering which can be driven by negligible interplane coupling.  The inset shows the same measurements to higher temperatures that indicate a chemical change, presumably at the surface, that generates ferromagnetic impurity. The inset also shows the susceptibility while cooling from the high temperature with a new superimposed FM component with a $T_{\rm c} \approx 560$ K. (c) Magnetization versus applied magnetic field at room temperature before and after heating the {\Ksub} sample used for the measurement shown in (b).}
\label{Fig:Susce1}
\end{figure}

The temperature dependence of the magnetic susceptibilities, $\chi\equiv M/H$ at an applied field of $H=1$ T along $a$ direction of single crystals of {\parent} and {\Ksub} are compared  in Figure \ref{Fig:Susce1}(a). It is evident that the {\Tn} for {\Ksub} is slightly higher than {\parent}. To better locate {\Tn}, we identify a peak in d$\chi (T)$/d$T$ versus $T$ as shown in the inset of Fig.~\ref{Fig:Susce1}(a) \cite{Curely1998}. We note that our $\chi$($T$) measurements, and the neutron diffraction results of both crystals described below, are consistent with previous reports of a C-type AFM order\cite{Liu2017}.

Two features in the susceptibility indicate that both samples retain 2D AFM correlations at temperatures above {\Tn}.  First, there is only a subtle signature at {\Tn} in the form of a small change in slope ({\Tn}  is  confirmed by neutron diffraction data, as discussed below).  Second, the susceptibility above {\Tn} does not display the characteristic paramagnetic behavior, namely, $\chi(T) \propto 1/T$ above the transition, but rather increases linearly with temperature above {\Tn} [$\chi(T) \propto T$], as shown in Figure \ref{Fig:Susce1}(b). The increase in susceptibility above  {\Tn} and the broad peak are common signatures  of 2D magnetic systems that develop large correlations prior to the 3D ordering, which can be driven by negligible interplane coupling.  The inset in Fig.\ \ref{Fig:Susce1}(b) shows the same measurements  to higher temperatures including during the cooling of the crystals that clearly show the emergence of a chemical change, presumably at the surface, that generates ferromagnetic impurity. While cooling from the high temperature a new superimposed FM component with a $T_{\rm c} \approx 560$ K is apparent. Figure\ \ref{Fig:Susce1}(c) shows magnetization {\it versus} applied magnetic field (hysteresis curve  with applied in the b-c plane) at room temperature of the same crystal on which the susceptibility shown in Fig.\ \ref{Fig:Susce1}(b) was measured on indicating the emergence of FM component after heating the sample.  Multiple cycles of heating/cooling the crystal show that the FM component does not increase after the second one (data not shown).

\begin{figure}
\includegraphics[width=2.4 in]{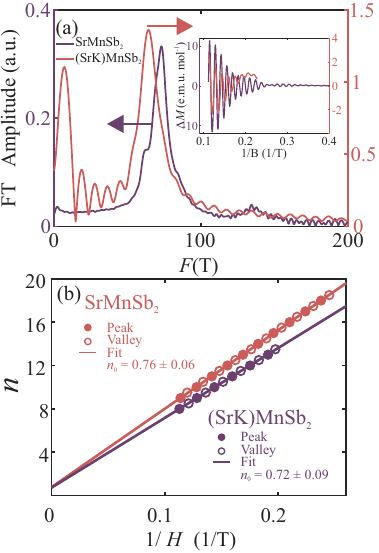}
\caption{(Color online) (a) Comparison of the FT of the oscillating $ M_a(B)$ for the parent {\parent} and the K-substituted {\Ksub}.  The inset shows $\Delta M_a(1/H)$ for both crystals as indicated.(b) Enumerating peak and valleys in (a) as a function of $1/H$ (assigning integer to minima). The intercepts at $n(1/H\rightarrow 0)=0.76(6)$ and $0.72(9)$ for both crystals as indicated.}
\label{Fig:magnetic_fft_comparison}
\end{figure}

To further characterize the parent and K-doped crystals, we measured the magnetization versus applied magnetic field along $a$ and in the $bc$ plane [$M_a(H)$ and $M_{bc}(H)$.  Whereas $M_{bc}(H)$ exhibits linear dependence on $H$  as is usual for applied field perpendicular to the moment direction in AFM crystals (not shown), the $M_a(H)$ exhibits strong de Haas van Alphen (dHvA) oscillations.  The inset in Fig.\ \ref{Fig:magnetic_fft_comparison}(a)  shows the oscillating part of the magnetization versus $1/H$  after subtracting an analytical function $M_a(H)=c_1{\rm erf}(c_2H)+c_3H+c_4$ from raw data with adjusted  parameters $c_i$. By performing a numerical Fourier transform (FT)  we obtain the spectral FT signals shown in Fig.\ \ref{Fig:magnetic_fft_comparison}(a).  The spectra shows similar features to those obtained in Ref. \cite{Liu2017,Liu2019} with a prominent peak at 73.1(2) T and a second peak at 137.0(5) T, not quite at the second harmonic position.   We note that for the K-substituted sample, the main peak shifts to lower fields from 73.1(2) T for the parent compound to 64(1) T. In addition, we consistently observe a low frequency peak at $\sim 7.3$ T, which can be an artifact due to the low quality of the signal and  due to the cutoff at large 1/H (K-substituted crystal are much smaller than the parent ones). 

Figure\ \ref{Fig:magnetic_fft_comparison}(b) shows the enumeration of peaks and valleys of $\Delta M_a(1/H)$  by assigning integers to minima.  The intercepts for both crystals are at $n(1/H\rightarrow 0)=0.76(6)$ and $0.72(9)$ for the {\parent} and {\Ksub}, respectively.  The value of the intercept relates to the Berry phase. It was reported that the intercept $n$ is 0.55 in Sr$_{1-y}$Mn$_{1-z}$Sb$_2$ with $y \sim 0.08$ and $z \sim 0.02$ through an analysis of Shubnikov-de Haas (SdH) oscillation, close to the expected value of 0.55 for a 2D or quasi-2D system with relativistic fermions\cite{Liu2017}. However, another transport analysis yields $n = 0.14$ for SrMnSb$_2$ with relatively low saturated magnetization moment, which suggests trivial topology\cite{Ramankutty2018}. The hole doping does not affect the $n$ value extracted from the analysis of dHvA oscillations, suggesting that both the parent and hole doped compounds share  similar topological characteristic, trivial or nontrivial. The intercept we observe in the dHvA oscillations [Fig.\ \ref{Fig:magnetic_fft_comparison}(b)], whether related to a nontrivial Berry phase, or not,  is still an open question.

\begin{figure}
\includegraphics[width=3.1 in]{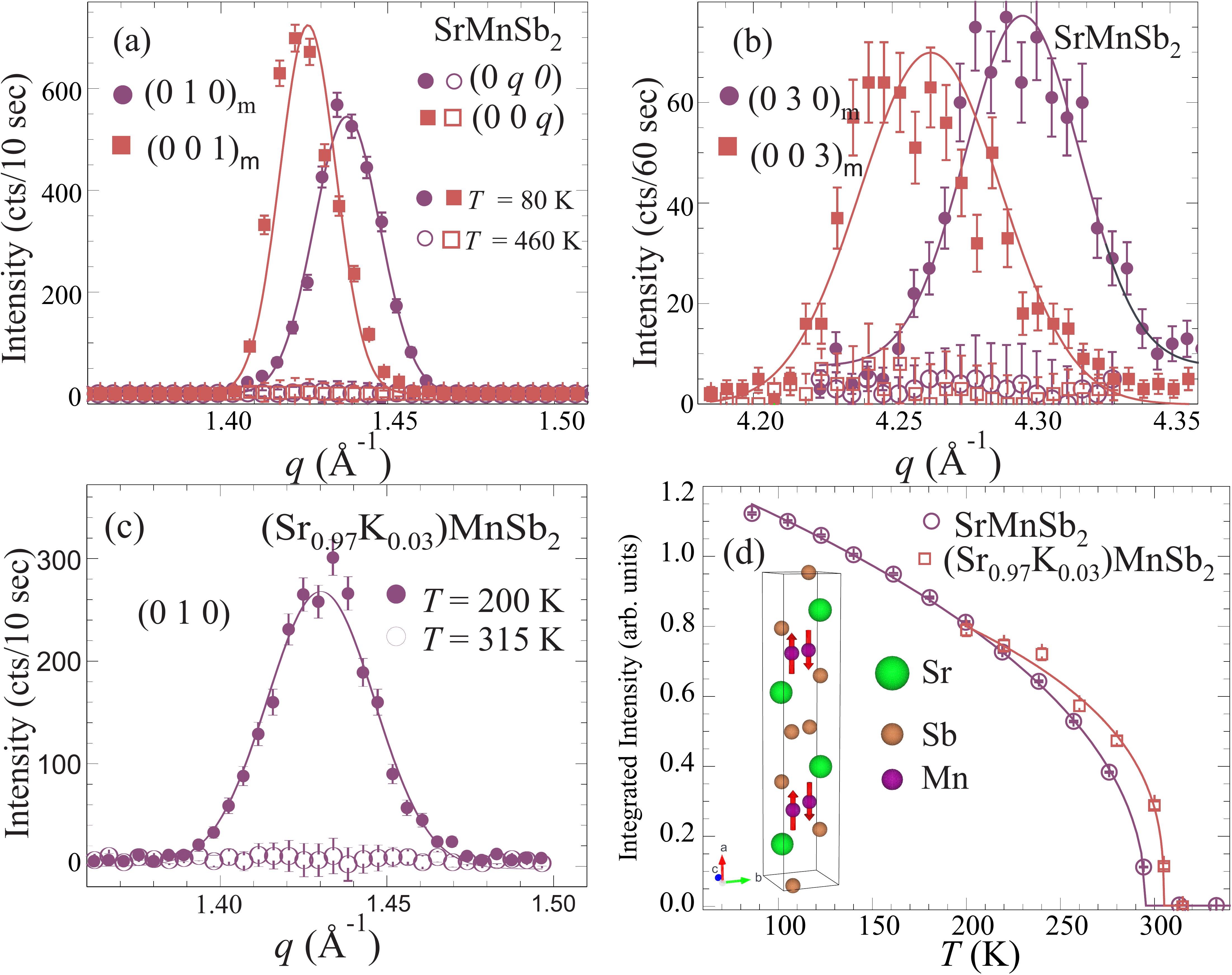}
\caption{(Color online) Intensity of magnetic Bragg reflections (a) (010) and (001) and (b) (030) and (003)  versus momentum transfer for  the stoichiometric single crystal SrMnSb$_2$ below and above {\Tn}. As is evident, the  crystal is to a large extent  not twinned. (c) Intensity of magnetic Bragg reflections (a) (010) for {\Ksub} below and above {\Tn} as indicated. (d) The intensities versus temperature of the magnetic (010) including a fit (solid line) to  a power law $I \propto (1-T/T_N)^{2\beta}$ yields $T_N = 297(2)$ and 305(4) K and $\beta = 0.227(2)$ and 0.173(3) for the pure and K-doped crystals  [(Sr$_{0.97}$K$_{0.03}$)MnSb$_2$], respectively. Inset illustrates atomic chemical and magnetic structure.}
\label{Fig:Peaks1}
\end{figure}

Figures\ \ref{Fig:Peaks1} (a) and (b) show the intensities of the (010)/(001)  and the (030)(003) AFM peaks respectively, at 80 and 460 K for   single crystal  SrMnSb$_2$  and  Fig.\ \ref{Fig:Peaks1}(c) shows the (010)/(001) AFM peaks  for the K-substituted (Sr$_{0.97}$K$_{0.03}$)MnSb$_2$. As shown, the distinct (un-split) planar orthorhombic peaks indicate a single crystal with almost no detectable orthorhombic twinning. Figure\ \ref{Fig:Peaks1}(d) shows the intensity of the (010) magnetic peaks below and above {\Tn}.  The magnetic peaks for both the parent and K-doped compounds are consistent with the magnetic structure previously reported \cite{Liu2017}. The intensities versus temperature of the (010) magnetic Bragg reflections with fit (solid lines) to a power law $I \propto (1-T/T_N)^{2\beta}$ yield  $T_N = 297(2)$ and 305(4) K and $\beta = 0.227(2)$ and 0.173(3) for the pure and K-substituted crystals, respectively. The inset in (d) illustrates the  chemical and magnetic structures. The increase in {\Tn} for the K-substituted crystal is in agreement with the determination of {\Tn} from the susceptibility measurements. Using  a few magnetic and nuclear Bragg reflections, we estimate the average staggered magnetic moment per Mn$^{2+}$ at 4.0(5) $\mu_B$ at 200 K  for both samples.  

\begin{figure}
\includegraphics[width=3.4 in]{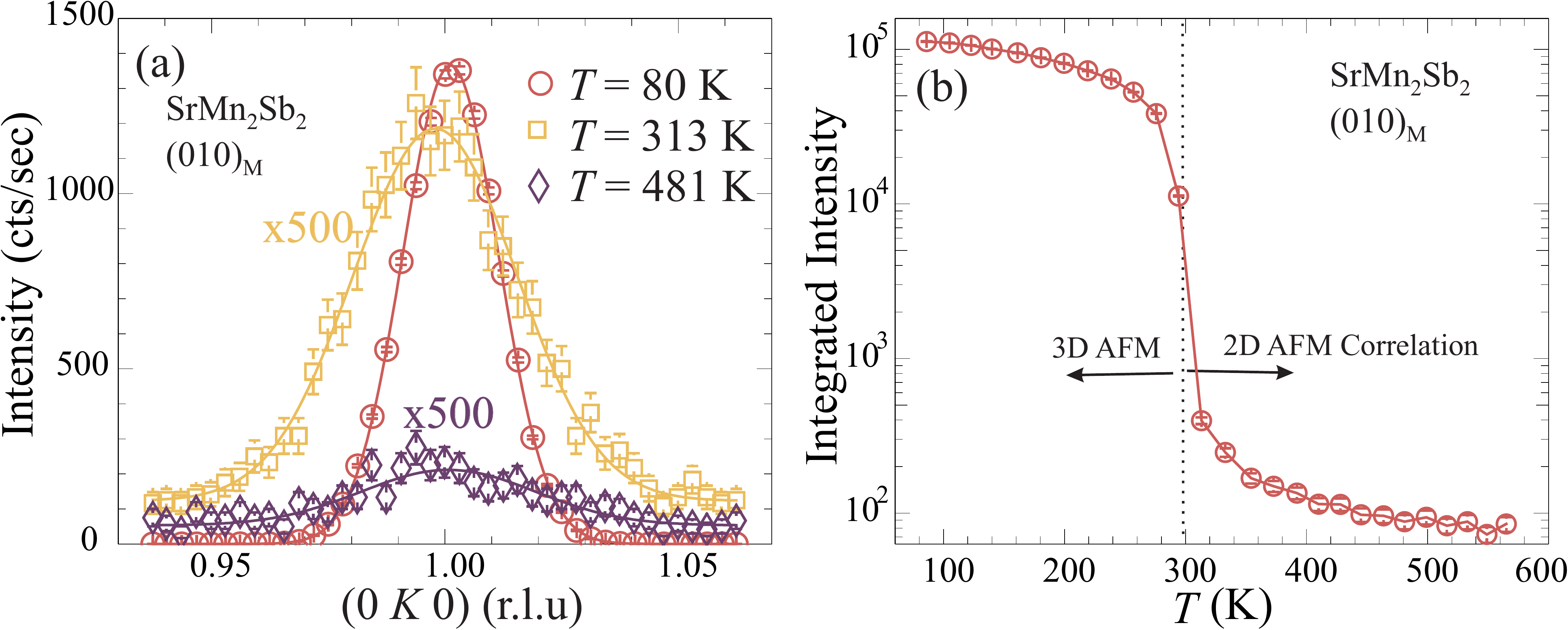}
\caption{(Color online)  (a) Magnetic Bragg reflection (010) at 80 K along with quasielastic scattering at 313 and 481 K (the intensities of peaks above {\Tn} are multiplied  by a factor of 500).  The FWHM of the peaks above {\Tn} suggests an upper limit for the average 2D correlations at 80 {\AA}. (b) Temperature dependence of the integrated intensity of the (010) peak  over the extended temperature range (on a semi-logarithmic scale) showing the persistence of magnetic scattering to the highest temperature measured with no indication of a transition up to  $\sim 575$ K.}
\label{Fig:Peaks2}
\end{figure}
An important finding in this study is the persistence of AFM correlations  (likely quasi-elastic) as observed for the magnetic (010) reflection (Fig.\ \ref{Fig:Peaks2}(a))  at temperatures significantly higher than {\Tn}.  These observations point to 2D correlations in the  Mn$Pn$ planes that develop at temperatures well above {\Tn}.  Figure\ \ref{Fig:Peaks2}  shows the  (010)  AFM peak at 80 K along with quasielastic scattering at 313 and 481 K (note that the intensities above {\Tn} are multiplied by a factor of 500).  The peaks above {\Tn} are much broader with a FWHM that sets an upper limit to the 2D correlations at $\sim80$ {\AA}. Temperature dependence of the integrated intensity of the (010) peak as a function of temperature is shown in Figure\ \ref{Fig:Peaks2} (b) over the extended temperature range (on a semi-logarithmic scale) indicating the persistence of magnetic scattering to the highest temperature measured with no indication of a transition up to  $\sim 575$ K.  This 2D scattering has been observed before for other 2D systems, in particular for the the 1111 type $A$Mn$Pn$O systems with planes that are weakly coupled magnetically \cite{Zhang2015,McGuire2016,Zhang2016,Liu2018}.

Our neutron diffraction and the temperature dependent susceptibility results  show that the nearly stoichiometric {\parent} and the K-doped compounds  do not exhibit unequivocal intrinsic spontaneous ferromagnetism.   Given that the potential interest in the {\parent}  is predicated on the development of a FM component in the bc-plane, careful consideration must be given to the presence of FM impurities, which may be present in the as-grown samples or may develop at high-temperature conditions, as shown in Fig.\ \ref{Fig:Susce1}(b) and as those reported recently\cite{Liu2019}. 

\section{Summary}

We report  neutron diffraction and magnetization measurements of single crystal {\Ksub} and compare them with those of the parent compound  {\parent}.  Our synthesis efforts indicate a very low limit in the substitution of K for Sr on order of a few percent.  We confirm that the K-substituted compound undergoes a C-type AFM transition below  {\Tn = 305(3)} K which is slightly higher than that  of the parent {\parent} for which {\Tn = 297(3)} K.  Both the parent and the K-substituted crystals show features in the high temperature susceptibility measurements (i.e., above {\Tn}) that are characteristic of quasi-2D systems evidence for 2D AFM  correlations.   Neutron diffraction measurements of the parent compound above {\Tn} are consistent with  the 2D-correlations that clearly persist to almost twice {\Tn}, and also confirm that the $Pnma$ symmetry is preserved with no detectible evidence for a FM phase  in  the 80 to 600 K temperature range. The persistence of such 2D correlations to high temperature is consistent with the layered structure, where magnetic Mn layers are well separated with interplanar coupling $J_a$ that is much weaker than the intralayer exchange constants $J_b$ or $J_c$, and consistent with spin-wave results of similar Mn layers in Mn-122 compounds.   We  caution that elevating the temperature of the crystals above $\sim 500$ K even under vacuum ($\sim 10^{-6}$ Torr),  as  in our susceptibility measurements, causes  a slight chemical change (presumably at the surface)  that creates a minute but detectible  FM impurity phase  with a $T_{\rm C} \approx 560$ K.  Magnetization measurements up to 9 T show strong dHvA oscillations for the {\Ksub} crystal  with Fourier Transform at  (${\rm FT}[M_a(1/H)]\simeq 64 $ T) compared to that of the {\parent} at ${\rm FT}[M_a(1/H)] = 73.1$ T.  Whereas the K-substitution affects the electronic properties as observed in dHvA oscillations, the hole doping of the system does not induce the desired itinerant ferromagnetism necessary to break time reversal symmetry. This may be due to the limited level of substitution of K for Sr in this system.

\acknowledgments

We thank N. S. Sangeetha for conducting the EDS measurements. This research was supported by the U.S. Department of Energy, Office of Basic Energy Sciences, Division of Materials Sciences and Engineering.  Ames Laboratory is operated for the U.S. Department of Energy by Iowa State University under Contract No.~DE-AC02-07CH11358. A portion of this research used resources at the High Flux Isotope Reactor, a DOE Office of Science User Facility operated by the Oak Ridge National Laboratory.


\end{document}